\documentclass{ws-ijmpe}
\usepackage[super,compress]{cite}
\begin{document}

\def\prl{{\em Phys. Rev. Lett. }}
\def\epja{{\em Eur. Phys. J.  A}}
\def\epl{{\em Europhys. Letts.}}
\def\prc{{\em Phys. Rev.  C }}
\def\prd{{\em Phys. Rev.  D }}
\def\jap{{\em J. Appl. Phys. }}
\def\ajp{{\em Am. J. Phys. }}
\def\nima{{\em Nucl. Instr. and Meth. Phys. A }}
\def\npa{{\em Nucl. Phys. A }}
\def\npb{{\em Nucl. Phys.  B }}
\def\njp{{\em New J.  Phys. }}
\def\epjc{{\em Eur. Phys. J.  C }}
\def\plb{{\em Phys. Letts. B }}
\def\phy{{\em Physics }}
\def\mpla{{\em Mod. Phys. Lett. A }}
\def\pr{{\em Phys. Rep. }}
\def\zpc{{\em Z. Phys. C }}
\def\zpa{{\em Z. Phys. A }}
\def\ppnp{{\em Prog. Part. Nucl. Phys. }}
\def\jpg{{\em J. Phys. G }}
\def\ijp{{\em Indian J. Phys.  }}
\def\ahep{{\em Adv. in High Eng. Phys. }}
\def\cpc{{\em Comput. Phys. Commun. }}
\def\app{{\em Acta Physica Pol. B }}
\def\aip{{\em AIP Conf. Proc. }}
\def\jhep{{\em J. High Energy Phy. }}
\def\psc{{\em Prog. Sci. Culture }}
\def\epjst{{\em Eur. Phys. J. ST }}
\def\snc{{\em Suppl. Nuovo Cimento }}
\def\sjnp{{\em Sov. J. Nucl. Phys. }}
\def\ptps{{\em Prog. Theor. Phys. Suppl. }}
\def\aipc{{\em AIP Conf. Proc. }}
\def\nt{{\em Nature }}

\markboth{Raghunath Sahoo}
{Transverse Energy and Charged Particle Production in Heavy-Ion Collisions: From RHIC to LHC}

\catchline{}{}{}{}{}

\title{Transverse Energy and Charged Particle Production in Heavy-Ion Collisions: From RHIC to LHC}

\author{Raghunath Sahoo\footnote{Corresponding Author: Raghunath.Sahoo@cern.ch}  ~and Aditya Nath Mishra}
\address{Indian Institute of Technology Indore\\
Indore, Madhya Pradesh-452017, India.\\}


\maketitle

\begin{history}
\received{Day Month Year}
\revised{Day Month Year}
\end{history}

\begin{abstract}
We study the charged particle and transverse energy production mechanism from AGS, SPS, RHIC to LHC energies in the framework of  nucleon and quark participants.  At RHIC and LHC energies, the number of nucleons-normalized charged particle and transverse energy density in pseudorapidity, which shows a monotonic rise with centrality, turns out to be an almost centrality independent scaling behaviour when normalized to the number of participant quarks. A universal function which is a combination of logarithmic and power-law, describes well the charged particle and transverse energy production both at nucleon and quark participant level for the whole range of collision energies. Energy dependent production mechanisms are discussed both for nucleonic and partonic level.  Predictions are made for the pseudorapidity densities of transverse energy, charged particle multiplicity and their ratio (the barometric observable, $\frac{dE_{\rm{T}}/d\eta}{dN_{\rm{ch}}/d\eta} ~\equiv \frac{E_{\rm{T}}}{N_{\rm{ch}}}$) at mid-rapidity for Pb+Pb collisions at $\sqrt{s_{\rm{NN}}}=5.5$ TeV. A comparison with models based on gluon saturation and statistical hadron gas is made for the energy dependence of $\frac{E_{\rm{T}}}{N_{\rm{ch}}}$.
\end{abstract}

\ccode{PACS Numbers: 24.85.+p}
\section{INTRODUCTION}	
Quantum Chromodynamics (QCD), the theory of strong interaction,
predicts a phase transition from normal hadronic matter  to a
deconfined state of quarks and gluons, called Quark Gluon Plasma
 (QGP) \cite{bjorken,evShuryak,evShuryak1,Karsch}. Such a
phase transition occurs at very high temperature and/or energy
density, and is also possible at lower temperature with very high baryo-chemical potential \cite{aoki}. It is possible to create these extreme conditions of high
temperature and/or energy density in the laboratory by colliding
heavy-ions at ultra-relativistic speeds. The properties of this new
phase of partonic matter is yet to be fully understood. 
The transverse energy and charged particle multiplicity are two key
observables in characterizing the bulk properties of the matter created in heavy-ion collisions.
Nuclei being extended objects, their collisions occur at various impact parameters and are characterized by collision centrality. The dependence of charged multiplicity and transverse energy on the collision geometry i.e. on centrality and collision energy is of paramount importance to understand the particle production
mechanism. The charge particle multiplicity provides important information on the initial entropy and its subsequent evolution in the hot and dense matter. This may give an insight to the partonic phase that might be created in heavy-ion collisions. The Large Hadron Collider (LHC) at CERN, provides a new domain of energy which is order of magnitude higher than the Relativistic Heavy Ion Collider at Brookhaven National Laboratory. Predictions of theoretical models that describe the particle production successfully at RHIC, vary almost a factor of 2 for  LHC energies \cite{armesto,abreu}. The LHC data have brought up new challenges in understanding the physics of heavy-ion collisions. 

 In this paper, we calculate the number of nucleon and quark participants in a nuclear overlap model in a way done previously to study the global properties of matter created in heavy-ion collisions \cite{voloshin,bm}, like transverse energy and charged particle multiplicity and the enhancement of multi-strange baryons in heavy-ion collisions \cite{me}. We study the centrality and collision energy dependence of charged particle and transverse energy production at mid-rapidity. After finding out a universality in particle production, we give the prediction for  $\frac{dN_{\rm{ch}}}{d\eta}$, $\frac{dE_{\rm{T}}}{d\eta}$ and the barometric observable, $\frac{dE_{\rm{T}}/d\eta}{dN_{\rm{ch}}/d\eta}(\equiv \frac{E_{\rm{T}}}{N_{\rm{ch}}})$ at mid-rapidity, for Pb+Pb collisions at $\sqrt{s_{\rm{NN}}}=5.5$ TeV. Possible physical reasons for the enhancement in $\frac{E_{\rm{T}}}{N_{\rm{ch}}}$ at LHC is discussed in the light of the proposed universality and gluon saturation picture. 
 \begin{figure}[ph]
\centerline{\psfig{file=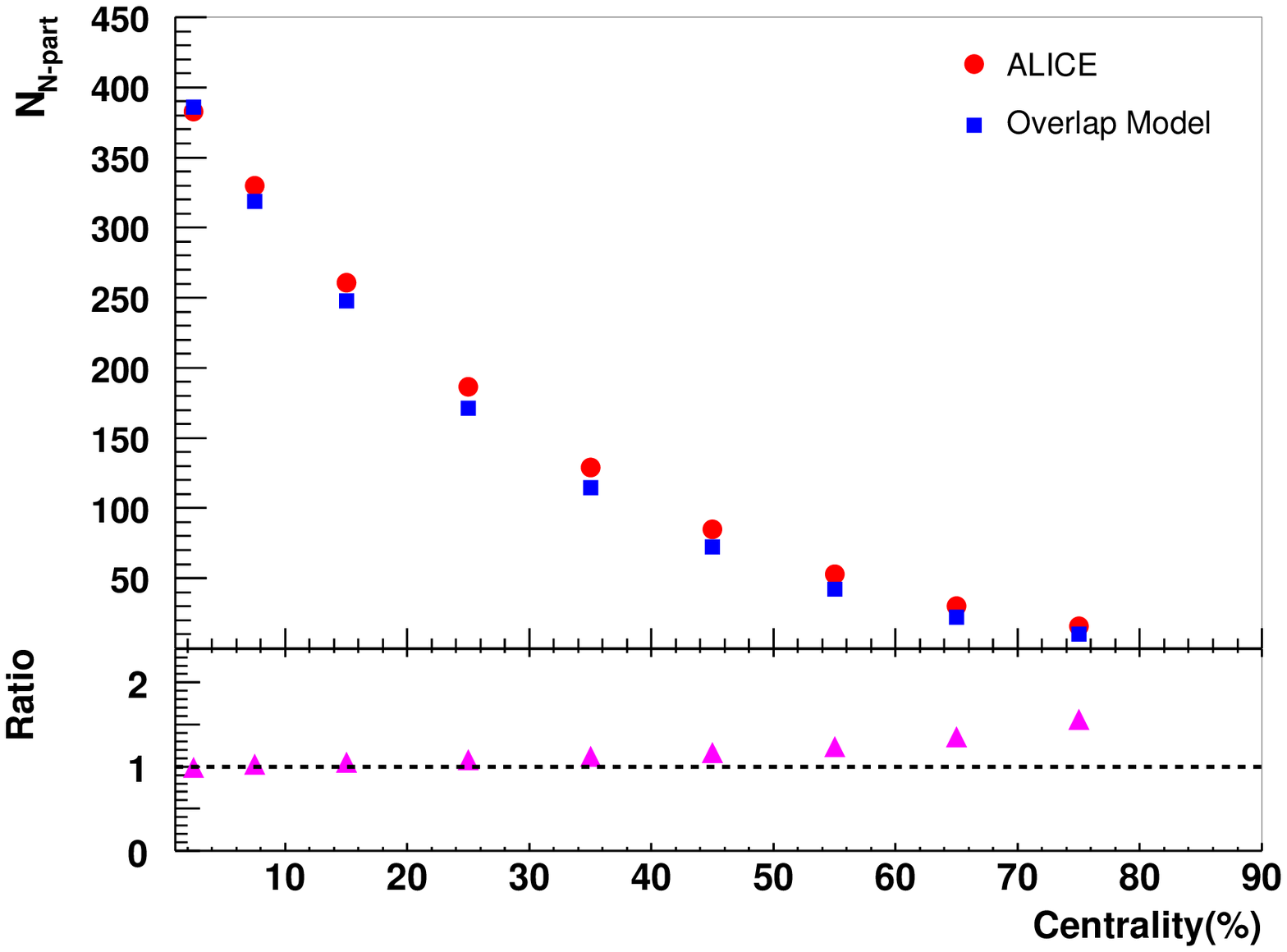,width=5.0in}}
\vspace*{8pt}
\caption{(Color online) The number of nucleon participants, $N_{\rm{N-part}}$ as
  a function of collision centrality for Pb+Pb collisions at $\sqrt{s_{\rm{NN}}}=2.76$ TeV from the 
  estimations by overlap model and by the ALICE collaboration \cite{aliceNpart}. The lower panel shows the ratio of both indicating the agreement of both the estimations. \protect\label{fig0}}
\end{figure} 
 
 \begin{figure}[ph]
\centerline{\psfig{file=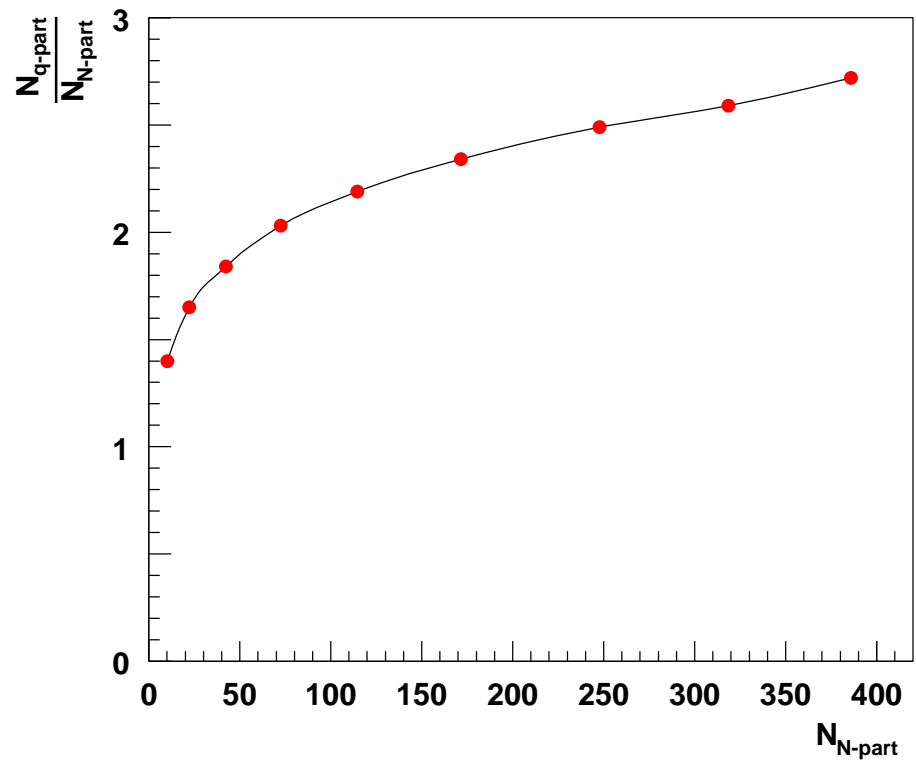,width=5.0in}}
\vspace*{8pt}
\caption{(Color online) The ratio of $N_{\rm{q-part}}/N_{\rm{N-part}}$ as
  a function of centrality for Pb+Pb collisions at $\sqrt{s_{\rm{NN}}}=2.76$ TeV from the 
  overlap model calculations. This increases monotonically 
  with collision centrality.\protect\label{nQByNp}}
\end{figure}

\begin{figure}[ph]
\centerline{\psfig{file=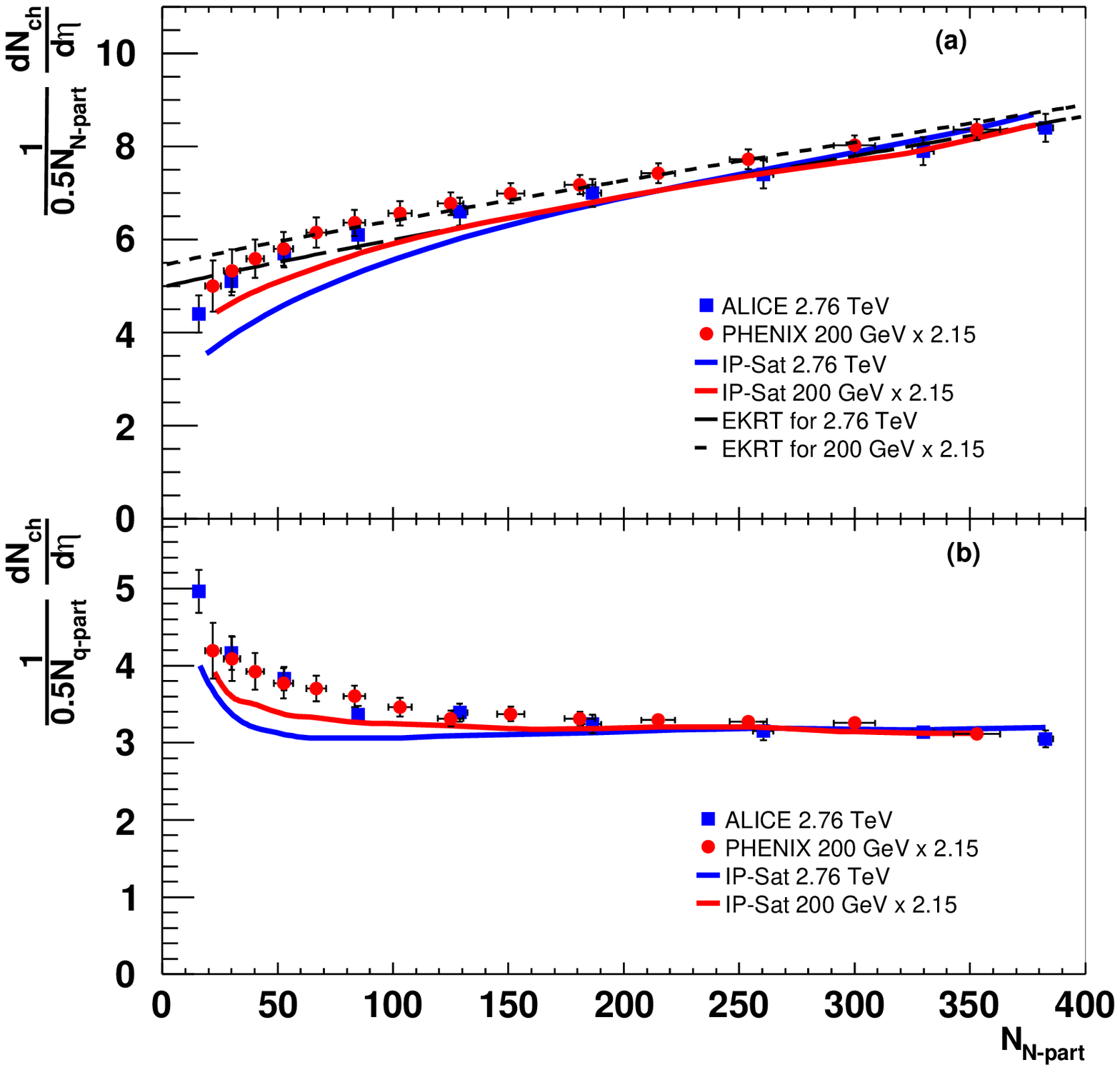,width=5.0in}}
\vspace*{8pt}
\caption{(Color online) (a) $0.5~N_{\rm{N-part}}$-normalized and (b)
  $0.5~N_{\rm{q-part}}$-normalized $\frac{dN_{\rm{ch}}}{d\eta}$ for
  Pb+Pb collisions at $\sqrt{s_{\rm{NN}}} =2.76$ TeV for LHC \cite{aliceNpart}
  and for Au+Au collisions at $\sqrt{s_{\rm{NN}}} =200$ GeV for RHIC
  \cite{phenixEt} as a function of collision centrality. The RHIC data are scaled to compare the shape of the distribution with LHC data, as is shown in the figure. Shown are the comparisons of IP-saturation model \cite{prithwish} and EKRT model \cite{ekrt}
predictions  for both the data sets.\protect\label{nChNpart}}
\end{figure}

\begin{figure}[ph]
\centerline{\psfig{file=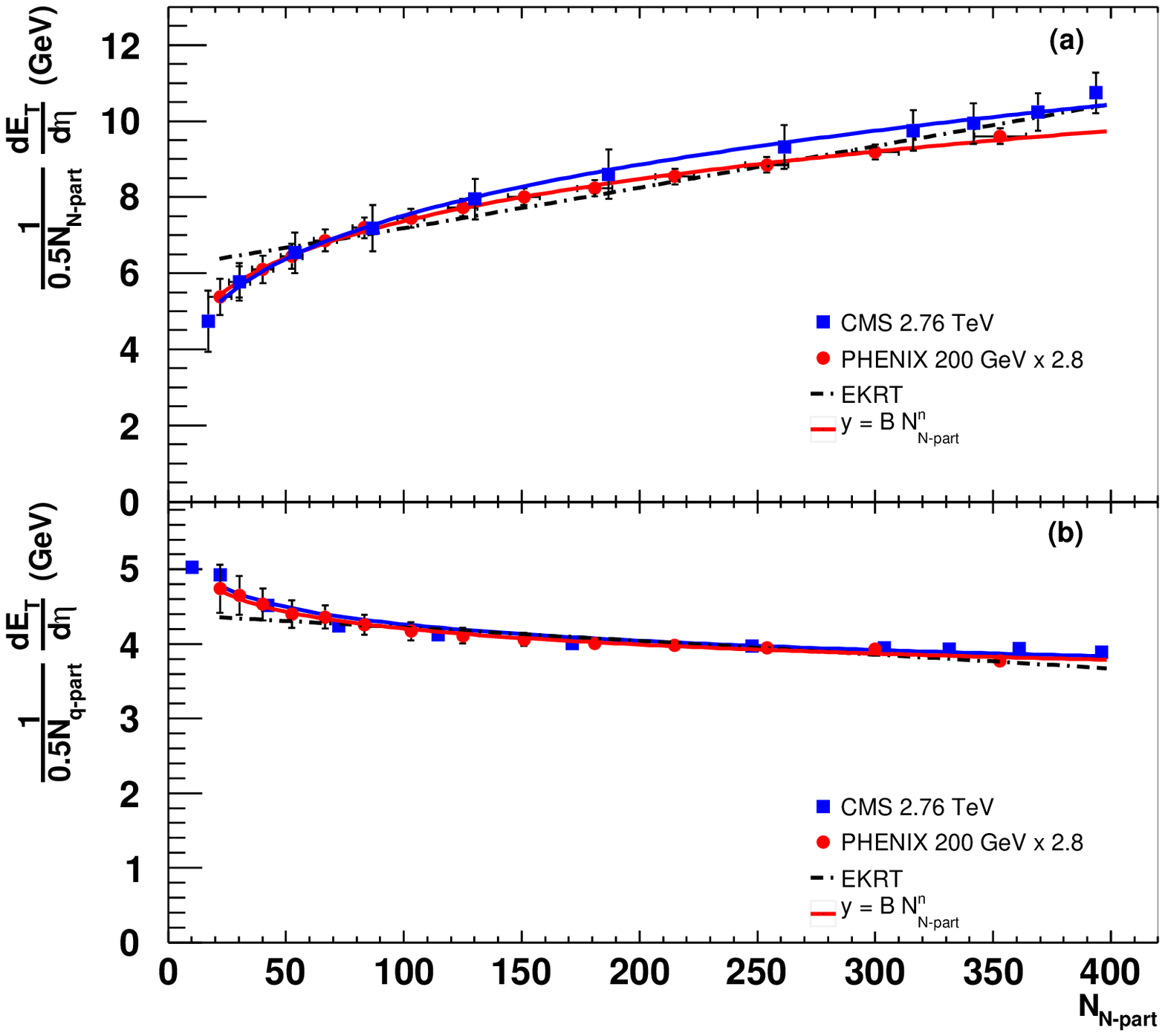,width=5.0in}}
\vspace*{8pt}
\caption{(Color online) (a) $0.5~N_{\rm{N-part}}$-normalized and (b)
  $0.5~N_{\rm{q-part}}$-normalized $\frac{dE_{\rm{T}}}{d\eta}$ for
  Pb+Pb collisions at $\sqrt{s_{\rm{NN}}} =2.76$ TeV for LHC
  \cite{cmsEt} and for Au+Au collisions at $\sqrt{s_{\rm{NN}}} =200$
  GeV for RHIC \cite{phenixEt} as a function of collision centrality. The RHIC data are scaled to compare the shape of the distribution with LHC data, as is shown in the figure.\protect\label{EtNpart}}
\end{figure}

\begin{figure}[ph]
\centerline{\psfig{file=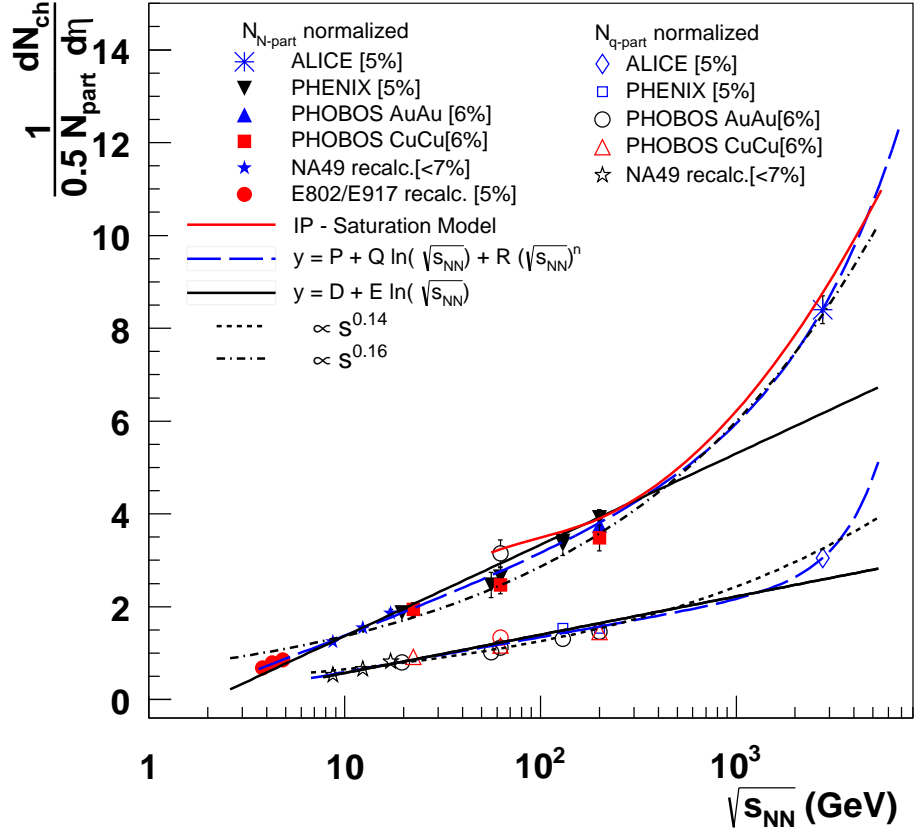,width=5.0in}}
\vspace*{8pt}
\caption{(Color online) The mid-rapidity $N_{\rm{N-part}}$ and $N_{\rm{q-part}}$ normalized $\frac{dN_{\rm{ch}}}{d\eta}$
 as a function of collision energy from lower AGS, SPS and RHIC
 \cite{phenixEt,phobos} to LHC energy \cite{aliceN}.
The collision data are compared with IP-saturation model \cite{prithwish} and 
hybrid function estimations.\protect\label{nChE}}
\end{figure}

\begin{figure}[ph]
\centerline{\psfig{file=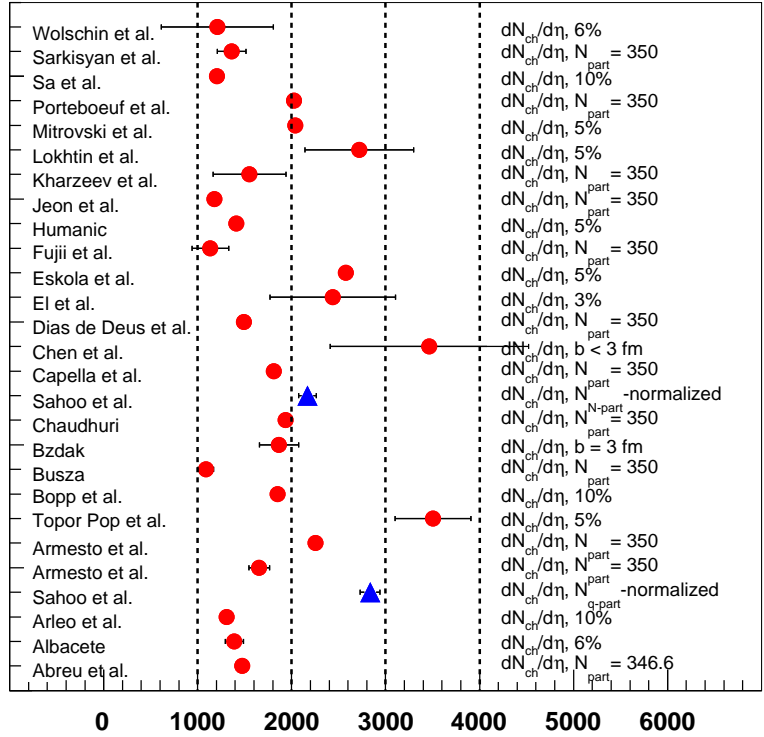,width=5.0in}}
\vspace*{8pt}
\caption{(Color online) Predictions for mid-rapidity
  $\frac{dN_{\rm{ch}}}{d\eta}$ for central Pb+Pb collisions at
  $\sqrt{s_{\rm{NN}}}= 5.5$ TeV at the LHC \cite{armesto}. On the left the names of the authors and on the right, the observables and the centrality definitions are shown. The error bars show the uncertainties in the predictions. Our predictions both from nucleon and quark participant normalizations are shown as filled triangles.\protect\label{predictions}}
\end{figure}

\begin{figure}[ph]
\centerline{\psfig{file=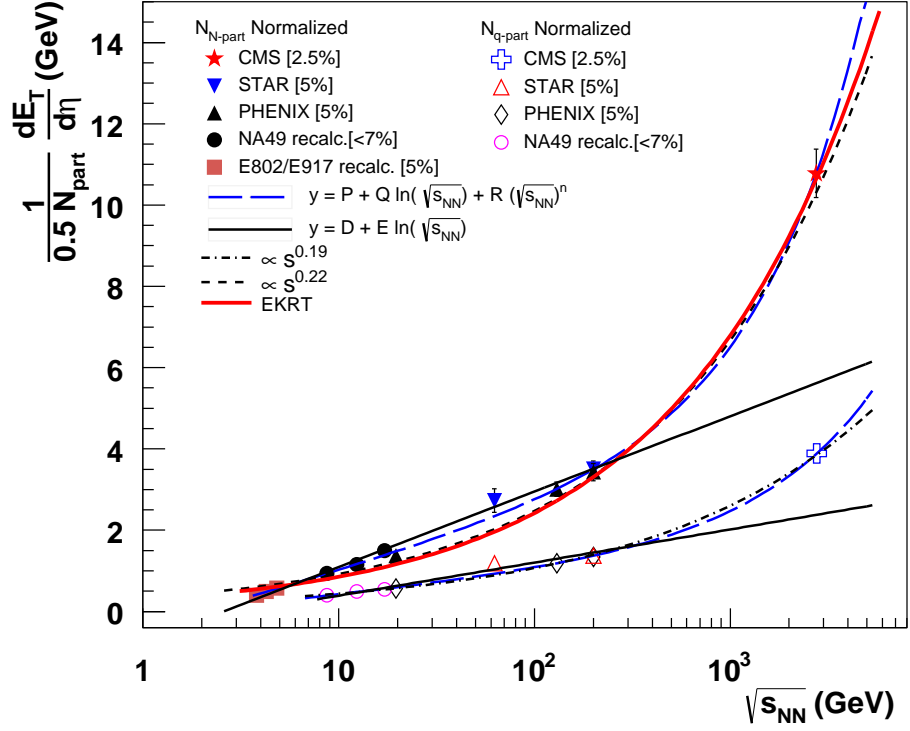,width=5.0in}}
\vspace*{8pt}
\caption{(Color online) The mid-rapidity $N_{\rm{N-part}}$ and $N_{\rm{q-part}}$ normalized $\frac{dE_{\rm{T}}}{d\eta}$
 as a function of collision energy from lower AGS, SPS and RHIC
 \cite{phenixEt,raghuThesis,raghuIJP} to LHC energy \cite{cmsEt}. 
 The collision data are compared with the EKRT model \cite{ekrt} 
 and the hybrid function predictions.\protect\label{etE}}
\end{figure}

\begin{figure}[ph]
\centerline{\psfig{file=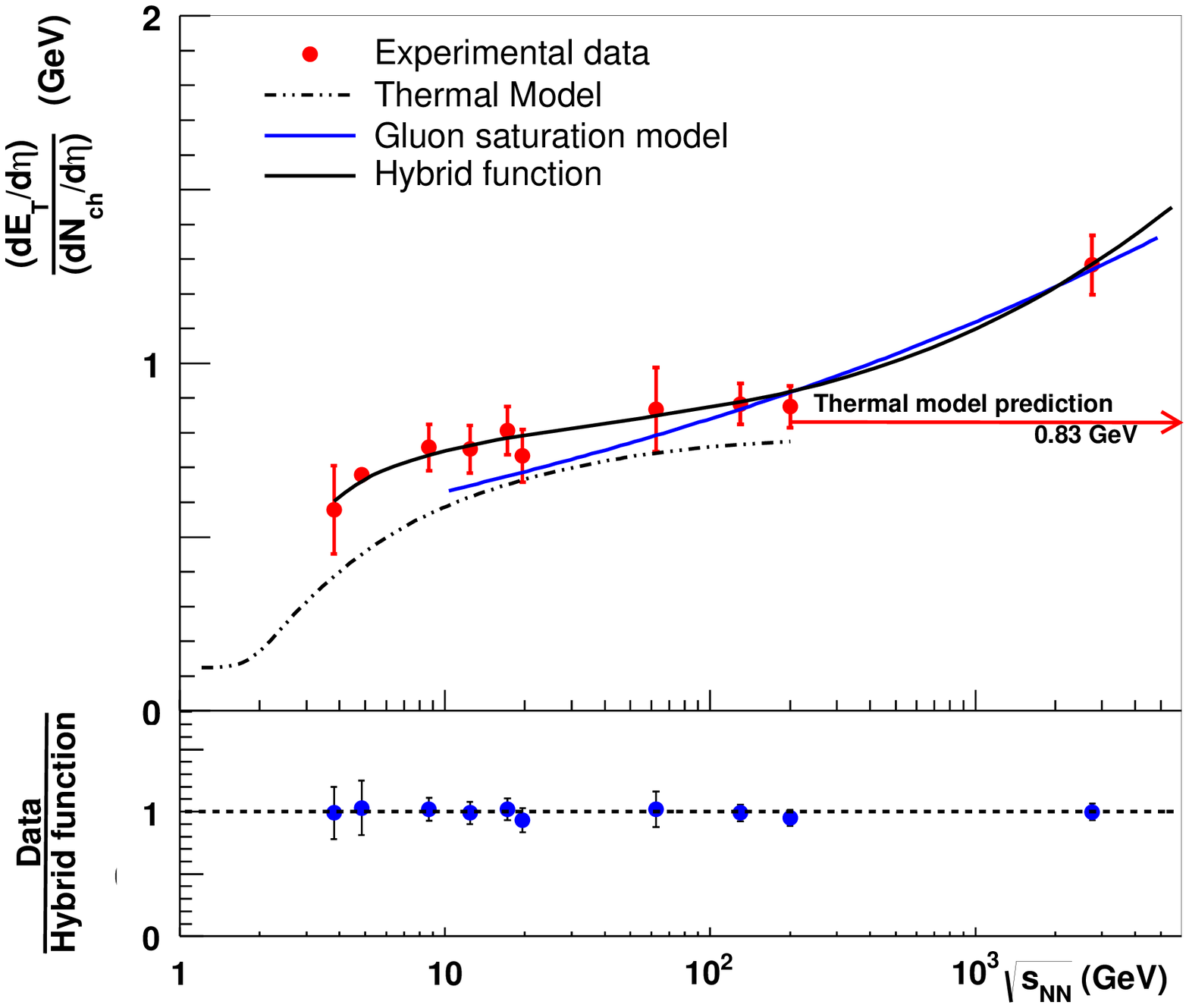,width=5.0in}}
\vspace*{8pt}
\caption{(Color online)  The ratio of $\frac{dE_T}{d\eta}$ and
  $\frac{dN_{ch}}{d\eta}$ at mid-rapidity, as a function of center of
  mass energy. Experimental data \cite{phenixEt,raghuThesis,raghuIJP,cmsEt} are compared to the
  predictions from thermal model \cite{rns}, gluon saturation model
  \cite{lappi,watt} and the estimations obtained in the framework of the hybrid model fitting to transverse energy and charged particle data.\protect\label{etEch}}
\end{figure}

\section{CALCULATION OF THE NUMBER OF PARTICIPANTS}
Here, both nuclei and nucleons are considered as superposition of
constituent or ``dressed'' quarks (partons or valons). There are three constituent 
quarks per nucleon. Baryons are composed of three and mesons are of two such quarks.
 The concept of constituent quarks is well known \cite{hwa,aniso,aniso11}
and was proposed in the realm of the discovery of constituent quark scaling of 
identified particle elliptic flow at RHIC \cite{starFlow}. The constituent quark
approach is successful in explaining many features of hadron-hadron, hadron-nucleus
and nucleus-nucleus collisions \cite{aniso1}. These include global properties like 
the charged particle and transverse energy density per participant pair 
\cite{voloshin,bm}. QCD calculations support the presence of 
three objects of size 0.1-0.3 {\it fm} inside a nucleon \cite{shuryak}. Further more it has been seen
that nucleus-nucleus collisions and p+p collisions have similar initial states
if the results are scaled by the number of constituent quark participants 
\cite{edward,edward1,rachid}. These observations also indicate that the particle production is
essentially controlled by number of constituent quark pairs participating in 
the collision. In a constituent quark picture, nucleon-nucleon ($NN$) collision looks like a collision of two light
nuclei with essentially one quark-quark ($qq$) pair interacting in the collision, leaving other
quarks as spectators. These quark spectators form hadrons in the nucleon
fragmentation region with a part of the entire nucleon energy being used for 
the particle production ($\sqrt{s_{\rm{qq}}}\sim \sqrt{s_{\rm{NN}}}/3$). Contrary, in $AA$ 
collisions, due to large nucleus size and the higher probability of quark-quark
interaction from same projectile and different target nucleons, more than one
quark per nucleon take part in the interaction. 

The calculations of the mean number of nucleon/quark participants are done using a Monte Carlo based implementation \cite{dm} of the nuclear overlap model \cite{eskola}. In the nuclear overlap model, the mean number of nucleon participants, $N_{N-part}$, in 
the collisions of a nucleus $A$ and a nucleus $B$ with impact parameter $b$ is given by
\begin{eqnarray}
N_{N-part, AB} = \int d^2s~T_A(\vec{s})\lbrace 1-\lbrack 1-\frac{\sigma_{NN}T_B(\vec{s}-
\vec{b})}{B}\rbrack ^B\rbrace \nonumber \\
+ \int d^2s~T_B(\vec{s})\{1-[1-\frac{\sigma_{NN}T_A(\vec{s}-\vec{b})}{A}]^A\},
\label{npart}
\end{eqnarray}
where $T(b)= \int_{-\infty}^{\infty} dz ~n_A(\sqrt{b^2+z^2})$ is the thickness function,
defined as the probability of having a $NN$ collision within the transverse area element $db$. 
$[1-\sigma_{NN}T_A(b)/A]^A$ is the probability for a nucleon to 
pass through the nucleus without any collision. $A$ and $B$ are the mass numbers of
two nuclei participating in the collision process. We use the Woods-Saxon nuclear density profile,

\begin{equation}
n_A(r) = \frac{n_0}{1+exp[(r-R)/d]},
\label{density}
\end{equation}

with parameters, the normal nuclear density $n_0 = 0.17 ~ fm^{-3}$, the nuclear
radius $R = (1.12 A^{1/3} - 0.86^{-1/3})~fm$  and the skin depth $d = 0.54~fm$.
The inelastic nucleon-nucleon cross section, {\it i.e.} $\sigma_{NN}$,
are $42\pm 3$ ~mb at $\sqrt{s_{\rm{NN}}} = 200 ~ GeV$ and $64\pm 5$ ~mb at $\sqrt{s_{\rm{NN}}} = 2.76$ ~ TeV \cite{X-sec}. 
In order to calculate the number of quark participants, $N_{\rm{q-part}}$, in nucleus-nucleus
collisions, the density for
quarks inside the nucleus is changed to three times that of the nucleon density
($n_0^q = 3 n_0= 0.51 ~ fm^{-3}$). Instead of nucleon-nucleon cross section, quark-quark 
cross section is used which is $4.67~mb$ and $7.1~ mb$ at
$\sqrt{s_{\rm{NN}}}=$ 200 ~GeV and 2.76 TeV respectively \cite{voloshin,bm,me,de}. 
For making predictions for Pb+Pb collisions at $\sqrt{s_{\rm{NN}}}=$ 5.5 ~TeV, we have used $\sigma_{NN} = 72 ~ mb$ \cite{totem} and $\sigma_{NN}/9 =8~ mb$ for $N_{\rm{N-part}}$ and $N_{\rm{q-part}}$ calculations, respectively.

\section{RESULTS AND DISCUSSION}
At lower center of mass energies, it has been found that the particle production 
scales with the number of participating nucleons, contrary to the case of high 
energies where hard processes dominate. Hard processes have much smaller 
cross-section than the soft processes. However, the number of binary collisions 
increase with increase in collision centrality faster than the number of 
participants, as a result the particle production per participant nucleon 
increases with increase in centrality. By using constituent quark approach, 
we are going to show how the particle production at higher energies depends on 
the participating quarks. It should be mentioned here that at energies higher to top
RHIC energy, gluons play a vital role in particle production. However, inclusion of 
the contributions from gluonic sources is beyond the scope of this paper. 
For these studies, we have taken the centrality dependence of 
pseudo-rapidity density of charge particles and transverse energy from top RHIC energy
to LHC $\sqrt{s_{\rm{NN}}}= 2.76$ TeV. To study the constituent quarks dependence of 
charged particle and transverse energy production, we need to estimate the number of 
participant quarks, which has been done in the framework of nuclear overlap model.\\

Since the quark participants are calculated in the framework of
nuclear overlap model, it is essential to check how good is our
estimation of number of participating nucleons in the collision.
In order to do that, the mean number of participating nucleons,
calculated in overlap model, are compared with the number estimated by
the ALICE experiment \cite{aliceNpart}. ALICE uses the Glauber MC technique for the estimation of the number of nucleon participants. We have found a reasonably good agreement   
with ALICE calculations excepting extreme peripheral collisions. This is shown in Figure \ref{fig0}. Solid circles represent the ALICE values and the
solid squares represent the overlap model calculations. The lower
panel of Figure \ref{fig0}, represents the ratio of ALICE values
and the overlap model values. 
We have then estimated the number of quark participants within the
prescription described in the previous sections. The ratio of quark
participants and nucleon participants is shown in Figure \ref{nQByNp} as a
function of nucleon participants, which is a measure of collision centrality. 
The ratio shows a sharp increase for $N_{N-part} \leq ~ 100$ and then shows 
a type of linear monotonic rise going from peripheral
to central collisions. 

\subsection{Centrality dependence of $\frac{dN_{\rm{ch}}}{{d\eta}}$}
Figure \ref{nChNpart} (a) shows the $0.5~N_{\rm{N-part}}$-normalized pseudorapidity density of charge particles at mid-rapidity for Pb+Pb collisions at $\sqrt{s_{\rm{NN}}} =2.76$ TeV for LHC and for Au+Au collisions at $\sqrt{s_{\rm{NN}}} =200$ GeV for RHIC, as a function of collision centrality. Henceforth, in this paper we would avoid mentioning "mid-rapidity" explicitly, unless otherwise required, as the data in general, are at mid-rapidity. A scaling factor of 2.15 has been multiplied with the RHIC data to make a direct comparison with that of LHC data. Going from peripheral to central collisions the production of charged particles show a rise, when normalized with number of participant pairs.  This rise in both the data sets are fitted with a function of the type:
\begin{equation}
\frac{1}{0.5N_{N-part}}\frac{dN_{ch}}{d\eta} = A N_{N-part}^n,
\label{cenEq}
\end{equation}
where, $A$ and $n$ are fitting parameters, the values of which are enlisted in Table \ref{ta1}, both for nucleon and quark participant sectors. The expectations from IP-saturation model \cite{prithwish} and the EKRT model \cite{ekrt}, which is based on initial state gluon saturation, are also shown for a direct comparison with data. In EKRT model the centrality dependence of charged particle multiplicity density in pseudorapidity is described by the function:
\begin{equation}
\frac{dN_{ch}}{d\eta}|_{b=0}= C ~\frac{2}{3}~ 1.16 (\frac{N_{N-part}}{2})^{0.92}(\sqrt{s})^{0.40}
\label{ekrtEq}
\end{equation}
Figure \ref{nChNpart} (b)
shows the $0.5~N_{\rm{q-part}}$-normalized $\frac{dN_{\rm{ch}}}{d\eta}$ for Pb+Pb collisions at $\sqrt{s_{\rm{NN}}} =2.76$ TeV for LHC and for Au+Au collisions at $\sqrt{s_{\rm{NN}}} =200$ GeV for RHIC as a function of collision centrality. A scaling factor of 2.15 has been multiplied with the RHIC data set to make a direct comparison with the corresponding LHC data. A similar function as is given by Eqn. \ref{cenEq} has been fitted to both the data sets. 
With top peripheral collisions as exceptions, where the calculation of nucleon and quark participants are not good, the $0.5~N_{\rm{q-part}}$-normalized $\frac{dN_{\rm{ch}}}{d\eta}$ both for RHIC and LHC data, show an almost centrality independent number of constituent quarks scaling behaviour. The comparison of both the above figures reveals that the particle production is better described in terms of constituent quarks towards more central collisions. This may be understood in line with the observation of very high initial energy densities being produced in central high-energy nuclear collisions \cite{phenixEt}. Hence, it seems more suitable to consider quarks as the sources of particle production than nucleons. 

\begin{table}[h]
\tbl{\bf{Power law fitting parameters for Fig. 3 and Fig. 4}}
{\begin{tabular}{@{}cccc@{}} \toprule
{\begin{tabular}{@{}ccccc@{}}
&&&$N_{\rm {N-part}}$-Normalization    \\
\colrule
&$\sqrt{s_{\rm {NN}}}$ (TeV) & A(B) &  n  \\
\colrule
 Fig3 & 0.2 & $2.74 \pm 0.24$ & $~0.183 \pm 0.017$ \\
        & 2.76 & $2.63 \pm 0.25$ & $~0.190 \pm 0.018$  \\ 
        \\
Fig4 & 0.2  & $2.9 \pm 0.20$ & $0.202 \pm 0.014$  \\
        & 2.76  & $2.5 \pm 0.40$ & $0.234 \pm 0.027$ \\ 
\botrule

\end{tabular}\label{ta1} }

{\begin{tabular}{@{}ccc@{}} 
&$N_{\rm{q-part}}$-Normalization\\

\colrule
 A(B) &  n \\
\colrule
 $5.38 \pm 0.40$ & $-0.091 \pm 0.014$\\
  $6.20 \pm 0.50$ & $-0.122 \pm 0.011$ \\ 
        \\
 $5.96 \pm 0.34$ & $-0.08 \pm 0.01$ \\
 $6.01 \pm 0.14$ & $-0.074 \pm 0.007$  \\
        \botrule 
       
        \end{tabular}\label{ta1} }
\end{tabular}\label{ta1} }
\end{table}


\subsection{Centrality dependence of $\frac{dE_{\rm{T}}}{{d\eta}}$}
The total transverse energy, which is measured on event-by-event basis, carries significant information regarding the explosiveness of the reaction and on characterizing the global properties of the produced system. The estimation of transverse energy provides model based calculation of the initial energy density produced in the collision. This in turn gives information regarding a possible formation of a partonic phase, when compared with that of lattice QCD prediction of critical energy density for the deconfinement transition \cite{Karsch}.  In nucleus-nucleus collisions, transverse energy is generated by the initial scattering of the partonic constituents of the incoming nuclei and by the rescattering of the produced partons and hadrons \cite{jacob,wang}. The transverse energy depends on the initial state of the collision and the viscosity of the partonic matter as it comes to the final state through interactions and expansion process \cite{peter}. Additionally it also depends on the entropy and temperature of the system \cite{cmsEt}. In Figure \ref{EtNpart} (a), $0.5~N_{\rm{N-part}}$-normalized $\frac{dE_{\rm{T}}}{d\eta}$
  for Pb+Pb collisions at $\sqrt{s_{\rm{NN}}} =2.76$ TeV for LHC and for Au+Au collisions at
 $\sqrt{s_{\rm{NN}}} =200$ GeV for RHIC are shown as a function of collision centrality. A scaling factor of 2.8 has been multiplied with both the $N_{\rm{N-part}}$ and $N_{\rm{q-part}}$-normalized RHIC data to make a direct comparison with the corresponding LHC data. Both the data sets show a functional trend which is governed by an equation,
\begin{equation}
\frac{1}{0.5N_{N-part}}\frac{dE_T}{d\eta} = B N_{N-part}^n,
\label{cenEqEt}
\end{equation}
where, $B$ and $n$ are fitting parameters. Here $B$ has the dimension of energy.  The values of the fitting parameters are given in Table \ref{ta1}, both for nucleon and quark participant sectors.
Figure \ref{EtNpart} (b) shows the $0.5~N_{\rm{q-part}}$-normalized $\frac{dE_{\rm{T}}}{d\eta}$ for Pb+Pb collisions at $\sqrt{s_{\rm{NN}}} =2.76$ TeV for LHC and for Au+Au collisions at $\sqrt{s_{\rm{NN}}} =200$ GeV for RHIC as a function of collision centrality.  A similar function as is given by Eqn. \ref{cenEqEt} has been fitted to both the data sets.  With top peripheral collisions as exceptions, the $0.5~N_{\rm{q-part}}$-normalized $\frac{dE_{T}}{d\eta}$ both for RHIC and LHC data within errors, show a nearly centrality independent number-of-constituent-quarks scaling behaviour. Similar observations have been made recently at RHIC energies \cite{phenix-2013}. The value of the power, {\it n} in the power law gets smaller (becomes negative) while going from $N_{\rm{N-part}}$ to $N_{\rm{q-part}}$-normalization, thereby making the spectra in $N_{\rm{q-part}}$ flatter compared to $N_{\rm{N-part}}$-normalization. The shapes of the centrality dependent  $N_{\rm{N-part}}$ or $N_{\rm{q-part}}$-normalized $\frac{dN_{\rm{ch}}}{d\eta}$ and $\frac{dE_{T}}{d\eta}$ show similar features for top RHIC and LHC energies. The relative enhancement of $N_{\rm{N-part}}$-normalized $\frac{dN_{\rm{ch}}}{d\eta}$ in A+A collisions with respect to that in $p+p$ collisions is the same at RHIC and LHC energies. This was seen in the framework of models based on multiple scattering and mutual boosting of saturation scales and shadowing \cite{pirner}. 

\subsection{Collision energy dependence of $\frac{dN_{\rm{ch}}}{{d\eta}}$}
In order to understand the collision energy dependence of charged particle and transverse energy production, we use the following fitting functions:
\begin {equation}
 \frac{1}{0.5N_{part}}\frac{dX}{d\eta} = D + E~ln~\sqrt{s_{NN}}
 \label{logarithm}
\end {equation}
\begin {equation}
 \frac{1}{0.5N_{part}}\frac{dX}{d\eta}  = F(s_{NN})^{n}
 \label{powerLaw}
\end {equation}
\begin {equation}
 \frac{1}{0.5N_{part}}\frac{dX}{d\eta}  = P + Q~ ln~\sqrt{s_{NN}} + R (\sqrt{s_{NN}})^n
 \label{hybrid}
\end {equation}
Here $X= N_{\rm {ch}}$ or $E_{\rm {T}}$ and $N_{\rm {part}}$ is either $N_{\rm {N-part}}$ or $N_{\rm {q-part}}$ depending on the case under discussion. The logarithmic and power law functions given by Eqs. \ref{logarithm} and \ref{powerLaw}, respectively, are motivated by the trend of the experimental data.
However, the form of the hybrid function given by Eq. \ref{hybrid} is motivated by the energy dependent behaviour of charged particle production and in addition, by the following results from the non-equilibrium statistical relativistic diffusion model (RDM) \cite{georg}. It is explicitly shown by the works of G. Wolschin {\it et al.}  \cite{georg1,georg2,georg3} that the RHIC and LHC multiplicity could be explained by the combination of a mid-rapidity gluonic source and a fragmentation source. Mid-rapidity gluonic sources predict a power-law type behaviour of charged particle multiplicity, whereas the fragmentation sources predict a logarithmic behaviour. Energies lower to that of RHIC, where particle multiplicity seems to obey a logarithmic behaviour, could be explained by fragmentation sources. The power law seems to take over after the top RHIC energy in describing the energy dependence of charged particle multiplicity and the transverse energy production. Additionally, it should be noted here that the trapezoidal approximation for the pseudorapidity distribution of charged particles \cite{busza,busza1,phobos}, which follows from the logarithmic behaviour with collision energy, seems to fail at LHC energies \cite{1304.0347}. 

In Figure \ref{nChE}, 
$\frac{dN_{\rm{ch}}}{d\eta}$ normalized to both $N_{\rm{N-part}}$ and $N_{\rm{q-part}}$ is shown 
as a function of collision energy. Up to top RHIC energy $N_{\rm{N-part}}$-normalized 
$\frac{dN_{\rm{ch}}}{d\eta}$ is well described by logarithmic function given
by Eq. \ref{logarithm} \cite{busza,busza1,phobos}. However, this function fails to describe the ALICE data for $2.76$ TeV. The estimation based on logarithmic function for the $N_{\rm{N-part}}$-normalized 
$\frac{dN_{\rm{ch}}}{d\eta}$ is $1180 \pm 22$ for $\sqrt{s_{\rm{NN}}} = 2.76$ TeV, which underestimates the data by $26\%$. A power law function (dash-dotted line) given by Eq. \ref{powerLaw}, discribes both RHIC and LHC heavy-ion data but overestimates lower-energy measurements. Contrary, the collision energy 
dependence of $\frac{dN_{\rm{ch}}}{d\eta}$ normalized by $N_{\rm{q-part}}$ is very well described 
by logarithmic function for the whole range of energies under discussion with little deviation towards LHC energies. In this case, the power-law gives a better description of data at the level of quark participants.
The power in the power law function decreases from $N_{\rm{N-part}}$ to $N_{\rm{q-part}}$
normalization, thereby going towards a flatter behaviour as a function of collision 
energy.  The predicted value of $\frac{dN_{\rm{ch}}}{d\eta}$ for Pb+Pb
collisions at $\sqrt{s_{\rm{NN}}} =$ 5.5 TeV based on the extrapolation of
power-law function fitted to $N_{\rm{q-part}}$ and
$N_{\rm{N-part}}$-normalization is around $2116 \pm 52$ and $2005 \pm 28$, respectively. On the other hand, the predicted value based on a logarithmic extrapolation of  $N_{\rm{N-part}}$-normalization data is $1310 \pm 22$. Looking at the low-energy and high-energy behaviours of charge particle production being well described by a logarithmic function and power-law functions, respectively, we have tried to fit a hybrid function (a combination of both) given by Eq. \ref{hybrid}. Both $N_{\rm{N-part}}$ to $N_{\rm{q-part}}$-normalized $\frac{dN_{\rm{ch}}}{d\eta}$ for the whole range of energies are very well described by this hybrid function, as is seen from Fig. \ref{nChE}. At the level of $N_{\rm{q-part}}$ and $N_{\rm{N-part}}$-normalization, the $\frac{dN_{\rm{ch}}}{d\eta}$ estimated by the hybrid function for Pb+Pb collisions at $\sqrt{s_{\rm{NN}}} =$ 2.76 TeV deviates from the experimental data by 0.06\% and 0.44\%, showing a reasonably good agreement with the data. The predicted values of  $\frac{dN_{\rm{ch}}}{d\eta}$ for Pb+Pb collisions at $\sqrt{s_{\rm{NN}}} =$ 5.5 TeV based on the extrapolation of the hybrid function fitted to $N_{\rm{q-part}}$ and $N_{\rm{N-part}}$-normalizations are around $2836 \pm 85$ and $2168 \pm 73$, respectively. The predictions from IP-saturation model for the top RHIC energy and higher, are also shown for a comparison with the corresponding experimental data. 

In Figure \ref{predictions}, the predictions for mid-rapidity $\frac{dN_{\rm{ch}}}{d\eta}$ for central Pb+Pb collisions at $\sqrt{s_{\rm{NN}}}= 5.5$ TeV at the LHC, are shown and compared with our predictions from the hybrid function at
nucleon and quark participant level. Other data are taken from Ref. \cite{armesto}. For central heavy-ion collisions at mid-rapidity, the charged particle multiplicity as a function of collision energy could be parametrized in the framework of the hybrid function, as 
\begin {equation}
\frac{dN_{\rm {ch}}}{d\eta}  = 0.5N_{\rm{N-part}}[P + Q~ ln~\sqrt{s_{NN}} + R (\sqrt{s_{NN}})^{n}]
 \label{hybridPara}
\end {equation}
where $P=-0.255\pm 0.100, Q =0.691\pm 0.141, R=0.0061\pm 0.0020~ GeV^{-1}$ and $n=0.789\pm 0.470$.


\begin{table}[h]
\tbl{\bf{Hybrid function fitting parameters for Fig. 5 and Fig. 7}}
{\begin{tabular}{@{}c@{}} 
\toprule

{\begin{tabular}{@{}cc@{}} 
$N_{\rm{N-part}}$-Normalization   & \\ \colrule

{\begin{tabular}{@{}cccccc@{}} 
& P &  Q & R & n\\
\colrule
 Fig5 & $-0.255 \pm 0.100$ & $0.691 \pm 0.141$  & $6.1\times 10^{-3} \pm 0.002$ & $0.789 \pm 0.470$ \\
 \\
 Fig7 &  $-0.416 \pm 0.121$ & $0.60 \pm 0.09$  & $8.9 \times 10^{-3} \pm 0.006$ & $0.832 \pm 0.211$ \\
\end{tabular}}
\end{tabular}}
\\
\colrule 

{\begin{tabular}{@{}cc@{}}
$N_{\rm {q-part}}$-Normalized    \\ 
{\begin{tabular}{@{}cccccc@{}} 
\colrule
 Fig5  &$-0.163 \pm 0.046$ & $0.327 \pm 0.005$  & $3.4 \times10^{-7} \pm 0.002$ & $2.108 \pm 0.285$ \\
 \\
 Fig7   & $-0.113 \pm 0.346$ & $0.218 \pm 0.217$  & $6.8 \times 10^{-3} \pm 0.045$ & $0.733 \pm 0.777$\\
\end{tabular}}

\\
\botrule 
\end{tabular}}
\end{tabular}}

\end{table}\label{ta2}


\begin{table}[h]
\tbl{\bf{Logarithmic function fitting parameters for Fig. 5 and Fig. 7}}
{\begin{tabular}{@{}cccc@{}} \toprule
{\begin{tabular}{@{}cccc@{}}
&&$N_{\rm {N-part}}$-Normalization   \\
\colrule
 & D &  E  \\
\colrule

Fig5 & $-0.60 \pm 0.07$ & $0.85 \pm 0.02$ \\
 \\
Fig7 & $-0.76 \pm 0.07$ & $0.81 \pm 0.02$ \\
 
\botrule

\end{tabular}\label{ta3} }

{\begin{tabular}{@{}ccc@{}} 
&$N_{\rm {q-part}}$-Normalization\\

\colrule
 D &  E \\
\colrule

 $-0.26 \pm 0.04$ & $0.36 \pm 0.01$ \\
 \\
 $-0.43 \pm 0.05$ & $0.35 \pm 0.02$\\ 

        \botrule 
       
        \end{tabular}\label{ta3} }
\end{tabular}\label{ta3} }
\end{table}
\begin{table}[h]
\tbl{\bf{Power law fitting parameters for Fig. 5 and Fig. 7}}
{\begin{tabular}{@{}cccc@{}} \toprule
{\begin{tabular}{@{}cccc@{}}
&&$N_{\rm {N-part}}$-Normalization    \\
\colrule
 & F &  n\\
\colrule

Fig5 &$0.65 \pm 0.02$ & $0.33 \pm 0.01$ \\
 \\
Fig7 &$0.34 \pm 0.02$ & $0.441 \pm 0.010$ \\ 
 
\botrule

\end{tabular}\label{ta4} }

{\begin{tabular}{@{}ccc@{}} 
&$N_{\rm {q-part}}$-Normalization\\

\colrule
 F &  n \\
\colrule
$0.34 \pm 0.01$ & $0.286 \pm 0.011$ \\
 \\
 $0.18 \pm 0.01$ & $0.386 \pm 0.007$ \\    

        \botrule 
       
        \end{tabular}\label{ta4} }
\end{tabular}\label{ta4} }
\end{table}


\subsection{Collision energy dependence of $\frac{dE_{\rm{T}}}{{d\eta}}$}
     A similar analysis is done for $\frac{dE_{T}}{d\eta}$-normalized to
$N_{\rm{N-part}}$ and $N_{\rm{q-part}}$ and is plotted as a function of
collision energy in Figure \ref{etE}. Up to top RHIC energy $N_{\rm{N-part}}$-normalized 
$\frac{dE_{\rm{T}}}{d\eta}$ is well described by logarithmic function given
by Eq. \ref{logarithm}. However, this fails to describe the ALICE data for $2.76$ TeV. The estimation of $\frac{dE_{\rm{T}}}{d\eta}$ based on logarithmic function for the $N_{\rm{N-part}}$-normalized $\frac{dE_{\rm{T}}}{d\eta}$ is $1114 \pm 24$ GeV, which underestimates the data by $47\%$ for $\sqrt{s_{\rm{NN}}}= 2.76$ TeV. A power law function (dash-dotted line) given by 
Eq. \ref{powerLaw}, discribes both RHIC and LHC heavy-ion data but underestimates lower-energy measurements. Contrary, the collision energy dependence of $\frac{dE_{\rm{T}}}{d\eta}$ normalized by $N_{\rm{q-part}}$ is very well described by logarithmic function up to top RHIC energy and shows considerable deviation at energies higher to top RHIC energy. The power law function given by Eq. \ref{powerLaw} describes the $N_{\rm{q-part}}$-normalized $\frac{dE_{T}}{d\eta}$ data quite well.
Like $\frac{dN_{\rm{ch}}}{d\eta}$, the power in the power law
function, decreases from $N_{\rm{N-part}}$ to $N_{\rm{q-part}}$
normalization, thereby going towards a flatter behaviour as a function of collision 
energy. The most interesting observation is that the power-law
motivated function better describes the transverse energy production
with collision energy compared to the logarithmic function.  The predicted value of $\frac{dE_{\rm{T}}}{d\eta}$ for Pb+Pb collisions at $\sqrt{s_{\rm{NN}}} =$ 5.5 TeV based on the extrapolation of
power-law function fitted to $N_{\rm{q-part}}$ and $N_{\rm{N-part}}$-normalized $\frac{dE_{T}}{d\eta}$ are around $2683 \pm 62$ GeV and $2760 \pm 50$ GeV, respectively. 
Like the case of charge particle production, looking at the low-energy and high-energy behaviours of the transverse energy production being well described by a logarithmic function and power-law functions, respectively, we have tried to fit the same hybrid function given by Eq. \ref{hybrid}. The $N_{\rm{N-part}}$-normalized $\frac{dE_{\rm{T}}}{d\eta}$ for the whole range of energies is very well described by this function, as is seen from Figure \ref{etE}. The $N_{\rm{q-part}}$-normalized $\frac{dE_{\rm{T}}}{d\eta}$ is also very well described by the hybrid function. Irrespective of nucleons or quarks are the sources of particle production, this hybrid function describes the charged particle and transverse energy production for all range of energies starting from lower AGS, SPS, RHIC to LHC. The estimations of $\frac{dE_{\rm{T}}}{d\eta}$ from the hybrid function fitting to the $N_{\rm{q-part}}$ and $N_{\rm{N-part}}$-normalized data for Pb+Pb collisions at $\sqrt{s_{\rm{NN}}} = 2.76$ TeV deviate from the CMS collaboration experimental data \cite{cmsEt} by 0.05\% and 1.18\%, respectively. The predicted value of  $\frac{dE_{\rm{T}}}{d\eta}$ for Pb+Pb collisions at $\sqrt{s_{\rm{NN}}} =$ 5.5 TeV based on the extrapolation of the hybrid function fitted to $N_{\rm{q-part}}$ and $N_{\rm {N-part}}$-normalizations are around $3056 \pm 44$ GeV and $3222 \pm 42$ GeV, respectively. Like charged particle multiplicity, for central heavy-ion collisions at mid-rapidity, the $\frac{dE_{\rm {T}}}{d\eta}$ as a function of collisions energy could be parametrized in the framework of the hybrid function, as 
\begin {equation}
\frac{dE_{\rm {T}}}{d\eta}  = 0.5N_{\rm{N-part}}[P + Q~ ln~\sqrt{s_{NN}} + R (\sqrt{s_{NN}})^{n}]
 \label{hybridPara1}
\end {equation}
where $P=-0.416\pm 0.121$ GeV, $Q =0.60\pm 0.09$ GeV, $R=0.0089\pm 0.0060$ and $n=0.832\pm 0.211$.
The transverse energy production as a function of energy in the framework of EKRT model, which is based on initial state gluon saturation \cite{ekrt}, is given by,
\begin {equation}
 \frac{dE_T}{d\eta} = 0.46~A^{0.92}(\sqrt{s})^{0.40}\lbrack 1-0.012~ln~A+0.061~ln{\sqrt{s}}\rbrack,
 \label{ekrtFn}
 \end {equation}
 where $A$ is the mass number of the colliding nuclei for a symmetric collision species.
 This takes into account a reduced initial number of scattering centers in the nuclear parton distribution functions. Here we have compared the EKRT model estimations for transverse energy for all collision energies under discussion. It is observed that the transverse energy production at higher energies is very well  described by EKRT model. However, EKRT model fails to describe the low energy data. This is an indication of gluon saturation effect at higher collision energies, which is absent at lower energies.

   To understand the deviation from logarithmic behaviour in a more qualitative way, one considers  a purely thermodynamic system in equilibrium. The entropy is proportional to the energy of the system, when the volume is assumed to be constant for non-expanding and homogeneous fireballs. This could be written as a Taylor expansion in $e^{\nu ~ln E} $, where $\nu$ is a constant and $E$ is the total energy of the system. Here $\frac{dN_{\rm{ch}}}{d\eta}/(0.5 N_{\rm{part}})$ represents a scaled-entropy \cite{BM-Jane} and at the freeze-out surface, where the volume of the fireball is fixed and we define a constant temperature of the system, this could be expressed as a function of collision energy up to second order in approximation, as:
  \begin{eqnarray}
 \frac{dN_{\rm{ch}}}{d\eta}/(0.5 N_{\rm{part}}) = e^{\nu~ln(\sqrt{s_{\rm{NN}}})}  \nonumber \\ 
  \simeq  \alpha + \beta~ ln\sqrt{s_{\rm{NN}}} +\gamma~ (ln\sqrt{s_{\rm{NN}}})^2
 \label{taylor}
 \end {eqnarray}
We have observed that both collision energy dependence of $\frac{dN_{\rm{ch}}}{d\eta}/(0.5 N_{\rm{part}})$ and $\frac{dE_{\rm{T}}}{d\eta}/(0.5 N_{\rm{part}})$ data up to top RHIC energy are explained by the above function. This may indicate that the produced system is thermalized and one uses equilibrium statistical mechanics to describe the system. However, LHC  data are very much underestimated by this function. The charged particle multiplicity at LHC has two components- thermalized soft component and a non-thermal hard component coming from jet fragmentation, which is almost more than $50\%$. 
The contribution of this non-thermal component to the final state entropy becomes significant at LHC, which may be the cause of this observed discrepancy.  This may also lead to a jump in $E_{\rm{T}}/N_{\rm{ch}}$ while going from top RHIC energy to LHC, where the statistical thermal model fails to explain the LHC data. This is discussed in the following section.

\subsection{Transverse Energy per Charged Particle and Freeze-out}
The ratio of pseudorapidity densities of transverse energy and number of charged particles at mid-rapidity i.e. $E_{\rm{T}}/N_{\rm{ch}}$ has been studied both experimentally \cite{phenixEt,cmsEt,starNpart} and phenomenologically \cite{rns,rns1,rns2,prorok} to understand the underlying particle production mechanism. This observable is known as global barometric measure of the internal pressure in the ultra-dense matter produced in heavy-ion collisions. This quantity depends on the initial state of the collision and the viscosity of the matter as it expands to its final state, when it is observed by the detectors. This observable when studied as a function of collision energy (as shown in Fig. \ref{etEch}),  shows two regions of interest. The first one from the lower SIS energies to SPS energies shows a steep increase of $E_{\rm{T}}/N_{\rm{ch}}$ values, thereby indicating that the mean energy of the system increases. In the second region, from SPS to top RHIC energy, $E_{\rm{T}}/N_{\rm{ch}}$ shows a very weak collision energy dependence, i.e. like a saturation behaviour. In this region the mean energy doesn't increase, whereas the collision energy increases. This may indicate that the increase in collision energy results in new particle production in this energy domain. This behaviour has been well described in the context of a statistical hadron gas model (SHGM) \cite{rns,rns1,rns2}. In the framework of SHGM, it has been predicted that $E_{\rm{T}}/N_{\rm{ch}}$  would saturate at energies higher to that of top RHIC energy with a limiting value of 0.83 GeV \cite{rns,rns1,rns2}.  This goes inline with Hagedorn's conjecture of a limiting temperature ($T_{\rm{H}} = 165$ MeV) for elementary and nuclear collisions at higher collision energies \cite{hagedorn,hagedorn1} and the subsequent observation of a saturation behaviour in the freeze-out temperature ($T_{\rm{fo}}$) in heavy-ion collisions \cite{tempSat}. The SHGM prediction of $E_{\rm{T}}/N_{\rm{ch}}$ at LHC collision energy of 2.76 TeV has been found to be inconsistent with the experimental observation of its value, which is $1.25 \pm 0.08$ GeV \cite{cmsEt}. This could be because of more high-$p_{\rm{T}}$ particle and jet production at LHC, which makes LHC measurements quite different from that of RHIC. The estimated values of $E_{\rm{T}}/N_{\rm{ch}}$ from the hybrid equation fitting to the corresponding $N_{\rm{q-part}}$-normalized data, for Pb+Pb collisions at 2.76 and 5.5 TeV collisions energies are  $1.33 \pm 0.05$ and $1.49 \pm 0.05$ GeV, respectively. Here we have taken the $N_{\rm{q-part}}$-normalized data and the corresponding fitting of the hybrid function, as it describes energy dependence of both charged particle and transverse energy production. This shows that the $E_{\rm{T}}/N_{\rm{ch}}$ shows a very strong dependence on collision energy beyond the top RHIC energy, thereby creating a third region in the $E_{\rm{T}}/N_{\rm{ch}}$ versus collision energy diagram. If the freeze-out occurs on a fixed isotherm for all energies, it is expected that the energy per particle will be the same. A comparison of $T_{\rm{fo}}$ and $E_{\rm{T}}/N_{\rm{ch}}$ versus collision energy indicates that from SPS to top RHIC energies the $T_{\rm{fo}}$ and the value of $E_{\rm{T}}/N_{\rm{ch}}$ are almost independent of collision energy. Beyond top RHIC energy, the value of $E_{\rm{T}}/N_{\rm{ch}}$ gets a jump, which indicates that the $T_{\rm{fo}}$ at LHC should be different than that is at RHIC. We have estimated the $T_{\rm{fo}}$ for the LHC energies from the SHGM parametrized equations used in Ref. \cite{tempSat}, which are given by 
\begin{equation}
T(\mu_B) = a -b \mu_B^2-c \mu_B^4,
\end{equation}  
where $a = 0.166 \pm 0.002~GeV$, $b= 0.139 \pm 0.016 ~GeV^{-1} $, and $c =0.053 \pm 0.021~ GeV^{-3}$ and
\begin{equation}
\mu_B(\sqrt{s}) = \frac{d}{1+e \sqrt{s}}
\end{equation}
where $d=1.308 \pm 0.028 ~ GeV$ and $e= 0.273 \pm 0.008~ GeV^{-1}$. The value of the freeze-out temperature, $T_{\rm{fo}}$, for RHIC 200 GeV, LHC 2.76 TeV and 5.5 TeV are around $166 \pm 2$ MeV. The  $T_{\rm{fo}}$ obtained from the SHGM estimation of fitting to experimentally obtained particle ratios at RHIC and LHC is $T_{\rm{fo}} = 164$ MeV, with $\mu_{\rm {B}}$ is $24$ MeV and $1$ MeV, respectively \cite{rhicTemp,lhcTemp}. The SHGM predictions for $T_{\rm{fo}}$ is consistent with its measurements at LHC. However, it should be mentioned here that the $\bar{p}/\pi^-$ ratio is sensitive to the determination of chemical freeze-out temperature. $T_{\rm{fo}}\equiv T_{\rm{chem}} = 164$ MeV, over predicts the value of $\bar{p}/\pi^-$ in hydrodynamical calculations \cite{pbarPi}, for the measurements at LHC energy\cite{lhcTemp,pbarPiLHC}, which is known as "proton puzzle" \cite{protonPuzz}. Inclusion of baryon anti-baryon $B-\bar{B}$ channels in phenomenological models \cite{vishnu} reduces the final state proton, anti-proton multiplicity and thus explains the proton puzzle of LHC.  On the other hand, the SHGM predicted value of $E_{\rm{T}}/N_{\rm{ch}}$, which is $0.83$ GeV is inconsistent with the corresponding measurements at LHC. The observation that $T_{\rm{fo}}$ doesn't change while going from RHIC to LHC energies, when $E_{\rm{T}}/N_{\rm{ch}}$ gets a jump, needs to be understood from thermodynamics view point. $T_{\rm{fo}}$ being independent of collision species and energy is a consequence of the fact that the system produced in heavy-ion collisions irrespective of the initial conditions comes to the same final state. $E_{\rm{T}}/N_{\rm{ch}}$, the mean energy per particle, getting enhanced beyond top RHIC energy, could be a consequence of LHC being dominated by high-$p_{\rm{T}}$ jets and as a barometric parameter, its increase implies, higher initial pressure being created in LHC collisions compared to that of RHIC. This also is reflected from the values of radial flow parameters measured at RHIC \cite{rhicBeta} and LHC \cite{lhcTemp}. Radial flow for central collisions at LHC is $10\%$ higher than the corresponding value at RHIC in the same $p_{\rm{T}}$ range \cite{lhcTemp}. At mid-rapidity, mean transverse mass ($<m_{\rm{T}}>$) is essentially the average energy of the particle ($E= m_{\rm{T}} cosh~y$) and is proportional to the temperature of the system in a thermodynamic picture. From RHIC \cite{rhicBeta,rhic-mT} to LHC \cite{lhcTemp} around $25 \%$ increase in $<m_{\rm{T}}>$ is observed, which could explain the observation of the jump in $E_{\rm{T}}/N_{\rm{ch}}$ from RHIC to LHC \cite{cmsEt}. However, there are other factors which may affect $<m_{\rm{T}}>$ and thus need to be considered for a proper interpretation of data \cite{BM-Jane}. 

In addition, this rise in $E_{\rm{T}}/N_{\rm{ch}}$ is expected in gluon saturation model due to the increase of the number of participating gluons with the increase of collision energy. At very high energy, the gluon creation and annihilation balance out leading to a saturation in gluon number. As gluon density does not have a linear collision energy dependence, the saturation scale depends non-linearly on the center of mass energy i.e. 
$E_{\rm{T}}/N_{\rm{ch}} = k (\sqrt{s})^{\lambda}$, where $k$ and $\lambda$ are constants \cite{lappi}. This function when fitted to the top RHIC and LHC data,  gives a good description of the energy dependent behaviour of  $E_{\rm{T}}/N_{\rm{ch}}$ and the value of $\lambda$ obtained from this fitting is found to be 0.12, which lies between $\lambda = 0.15$ (IP-sat model prediction) and $\lambda = 0.11$ (b-CGC model estimation) \cite{lappi,watt}. Hence, gluon saturation picture at LHC energies may also explain the observed jump in $E_{\rm{T}}/N_{\rm{ch}}$ \cite{peter}.

\section{SUMMARY AND CONCLUSION}

In this paper, we study the charged particle and transverse energy production as a function of collision centrality and collision energy, in the framework of nucleon and quark participants estimated using a Monte Carlo based implementation of nuclear overlap model. The ratio of number of quark participants and number of nucleon participants shows a non-linear increase with collision centrality for a particular collision energy. The centrality dependence of $N_{\rm{N-part}}$-normalized $\frac{dN_{\rm{ch}}}{d\eta}$ both for Au+Au collisions at $\sqrt{s_{\rm{NN}}}=200$ GeV and Pb+Pb collisions at $\sqrt{s_{\rm{NN}}}=2.76$ TeV, which show a increase with centrality, seems to show a nearly centrality independent scaling behaviour when normalized to $N_{\rm{q-part}}$. This gives the indication of partonic activities at RHIC and LHC energies. Similar is the observation for $\frac{dE_{\rm{T}}}{d\eta}$ both for the RHIC and LHC data. 

The charged particle production as a function of collision energy has been studied both in the nucleon and quark participant frameworks. In contrast to the nucleon participants where the logarithmic function fails to describe the measurements up to the LHC energies, in the quark participant sector, the logarithmic function describes the data well  up to 2.76 TeV.
 A power-law function however, describes the high energy data for mid-rapidity $\frac{dN_{\rm{ch}}}{d\eta}$ with a nice agreement with LHC 2.76 TeV measurement. On the other hand, it underestimates the low energy data. A hybrid function, which is a combination of logarithmic and power-law in collision energy, describes the data for $0.5~N_{\rm{part}}$-normalized $\frac{dN_{\rm{ch}}}{d\eta}$ and $\frac{dE_{\rm{T}}}{d\eta}$ at all available energies both for the nucleon and quark participants, indicating a universality in multi-particle production. We give the predictions for $\frac{dN_{\rm{ch}}}{d\eta}$ for Pb+Pb collisions at $\sqrt{s_{\rm{NN}}}=5.5$ TeV from this universal hybrid function. On the other hand, the EKRT model based on a initial state gluon saturation and the power-law function better describes the high energy behaviour of transverse energy production. This points to gluon saturation effects at energies higher to RHIC. We have estimated the value of $\frac{dE_{\rm{T}}}{d\eta}$ for Pb+Pb collisions at $\sqrt{s_{\rm{NN}}} =2.76$ TeV by using this hybrid model which agrees with the corresponding LHC measurements. Based on this, we give a prediction of $\frac{dE_{\rm{T}}}{d\eta}$ for the Pb+Pb collisions at $\sqrt{s_{\rm{NN}}} =5.5$ TeV. The value of $E_{\rm{T}}/N_{\rm{ch}}$ for Pb+Pb collisions at $\sqrt{s_{\rm{NN}}}=2.76$ TeV estimated by the hybrid function agrees with the corresponding measurement by CMS collaboration at LHC and reproduces the jump from RHIC to LHC, in its energy dependence behaviour. This may indicate a higher initial pressure of the ultra-dense matter at LHC compared to that is formed in RHIC energies. In this framework, we give the prediction for this barometric observable for the future LHC Pb+Pb collisions at $\sqrt{s_{\rm{NN}}}=5.5$ TeV.  In conclusion, the charged particle and the transverse energy measurements  are well described by the hybrid function for all available energies and collision species in heavy-ion collisions indicating a universality in particle production. 
 
 \vspace{0.5cm}
\noindent
\textbf {Acknowledgements}\\
Authors would like to thank Nirbhay K. Behera for fine tuning the nuclear overlap model code and Prof. B. Nandi for very exciting discussions. We are very much grateful to Prithwish Tribedy for providing us the IP-saturation data and for the stimulating discussions related to gluon saturation at high energy.

\end{document}